\documentstyle[aps, epsf, amsbsy]{revtex}
\begin{document}
\draft

\title{On Reduction of Critical Velocity in a 
   Model of Superfluid Bose-gas with Boundary Interactions} 

\author{D. Roubtsov,  Y. L\'epine}
\address{D\'epartement de Physique,
Groupe de Recherche en Physique et Technologie des Couches Minces,
Universit\' e de Montr\' eal,
C.P. 6128, succ. Centre-ville, Montreal, PQ, H3C\,3J7, Canada, 
e-mail: roubtsod@physcn.umontreal.ca }
\date{today}
\maketitle
\begin{abstract}
 
The existence of superfluidity in a 3D Bose-gas  
can depend on boundary interactions with channel 
walls. We study a simple model where the dilute moving Bose-gas  
interacts 
with the walls via hard-core repulsion. Special boundary excitations  
are introduced, and their excitation spectrum is calculated within a 
semiclassical approximation. It turns out that the state of the
moving Bose-gas is unstable with respect to the creation of these boundary
excitations in the system gas\,$+$\,walls, i.e. the critical   
velocity vanishes in the semiclassical (Bogoliubov) approximation. 
We discuss how 
a condensate wave function, the boundary excitation spectrum and, hence, the value of the critical velocity can change in more realistic
models, in which ``smooth'' attractive interaction between the gas and walls
is taken into account.
Such a surface mode could exist in ``soft matter'' containers 
with flexible walls.

\vspace*{0.5cm}

{\it Keywords}:  Nonideal Bose-gas, Boundary 
Interactions, Surface Excitations, Critical Velocity, Su-

perfluidity. 
\end{abstract}
\pacs{PACS numbers: 05.30.Jp, 64.60.Ht, 67.40.Db}

\section{Introduction}

Recent experiments on Bose-Einstein Condensation (BEC) in magnetically 
trapped gases \cite{Anderson},\cite{Davis}, 
and excitons in semiconductor crystals and nanostructures 
\cite{Lin},\cite{Mys},\cite{Butov}
have made the subject of
BEC more vital, more interdisciplinary.
Many new questions appeared naturally
as the understanding of the process 
of BEC progressed \cite{Data}. 
However, some ``old'' problems -- such as the kinetics of BEC, the nature of 
superfluidity, the critical velocity problem, etc. -- 
still remain the subject under consideration \cite{BoseCondensation}, 
(especially for 
new physical objects where the state of BEC was recently demonstrated 
\cite{Data}).      
Although the critical velocities are one of the first difficulties to be 
encountered in the study of superfluidity, they are still the least
understood aspect in the theory \cite{Huang},\cite{Rokhar},\cite{Zaremba}.

This article is motivated mainly by the problem of critical 
velocity(-ies) in exciton superfluidity. It has been found 
experimentally \cite{Mys},\cite{Benson},\cite{Mys1} that  
a cloud of condensed excitons moves through a 
crystal with some constant velocity and some characteristic shape of
the density profile. Several theoretical explanations of this anomalous
transport have been put forward \cite{Hanamura},\cite{Tichodeev}, 
\cite{exciton}. 
In spite of the fact that these explanations are based on different
assumptions, there are several common ideas in the background of
all these theories. For instance, 
 it is the notion that interaction with a lattice is very important 
(if not to say crucial) in the BEC of excitons 
\cite{exciton},\cite{Baym},\cite{Imam}. 

However, there are many outstanding questions that
remain the subject of discussion \cite{discussion}.     
One of such questions, for example, is the superfluid nature of 
exciton anomalous 
transport. In fact, it is not clear how the exciton condensate ``feels'' the 
boundary of the crystal (via interaction with surface phonons, e.g.,)
or the impurities and other lattice imperfections that can bound an 
exciton. Generally speaking, 
clarification of the role of these friction sources may be essential for
the understanding of the exciton superfluidity and the nature of critical 
velocities.

We approach the critical velocity problem by working out  
a simple model in which dilute 3D Bose-gas moves in a channel and interacts
with the walls of this channel. 
The walls are modeled as two 3D solid bodies with well-defined boundaries.
Although we take into account
repulsive interaction between the particles of the gas, the 
proposed model 
cannot describe, for example, the superfluid He, which is a Bose liquid 
with strong interparticle interaction. Yet this is not 
the aim of this article.   
The main goal of this study is to explore the space of man{\ae}uvre
appearing in the framework of the well known simple models, such as the 
weakly nonideal Bose-gas, if we switch on the gas-wall boundary 
interaction.  

We show that the existence of the repulsive 
interactions
between the Bose-gas and the channel walls leads to the essential
reduction of the critical velocity of the superflow.
The finiteness of the Landau critical velocity
 for the bulk (i.e. Bose-gas) excitations 
turns out
not to be a sufficient condition of superfluidity.
Note that in the present work we investigate the boundary excitations.
Although the breakdown of the superfluidity
is assumed to be accompanied by vortex emission, we
leave for future studies the questions of the vortex formation and their 
dynamics
in the case when interaction with walls is taken into account.


\section{Critical Velocity Problem  }

A closed system cannot undergo an inner macroscopic motion in thermodynamic
equilibrium. Once such a motion is present, the system must
evolve toward an
equilibrium state. However, unless this transition is kinematically 
prohibited,
(i.e. incompatible with conservation laws), the macroscopic motion is 
sustained. Such is the case with  a small object moving without 
any viscous drag in stationary
superfluid \cite{tilley},\cite{Huang1}.
This  object is assumed to have no inner degrees of freedom,
so that its momentum and energy depend only on the 
velocity $ {\bf v} $ of the center of mass.

If the conservation laws for the creation
of an excitation with the energy $ \epsilon_{\rm gas}(k) $ and 
the momentum $ {\bf k} $
in a fluid (gas)
lead to \cite{tilley}-\cite{body} 
\begin{equation}
\epsilon_{\rm gas}(k)>0,\,
{\rm in\,the\,object\,reference\,frame}, 
\label{criterion}
\end{equation}
the particle will continue to move without any experience of drag forces.
Condition (\ref{criterion}) is known as the Landau criterion of superfluidity.
In fact, Eq. (\ref{criterion}) holds that  \cite{tilley}-\cite{body}
\begin{equation}
v<v_{\bf L}=\left(\epsilon_{\rm gas}(k)/k\right)_{\rm min} .
\label{velocity}
\end{equation}
Here $v_{\bf L}$ is the Landau critical velocity.

Formula (\ref{velocity}) is in agreement with experiments performed
with the semimicroscopic objects moving in the liquid helium \cite{body}.

Formula (\ref{criterion}), (taken in the channel reference frame),  is 
employed as a criterion of Bose-gas superfluid 
flow  in channels. In that case, the channel walls are regarded as a
massive macroscopic body in the above consideration (i.e. the walls act 
as some source of perturbations on 
the gas flow), and the final result is 
formulated in the form  (\ref{velocity}).

On experiments with liquid helium flow, 
however, the registered values of the critical
velocities turn out to be much smaller than $ v_{\bf L} $.
Moreover, the critical velocity depends on the channel dimensions 
\cite{putterman}.
The fact that the liquid superfluid helium could not be treated as a dilute 
Bose-gas is believed to be the main reason for this discrepancy.
It is generally assumed that the critical velocities are related 
to the appearance
of quantized vortex lines in the superfluid.
The Landau criterion (\ref{criterion}) applied to the vortex excitations
\cite{onsager} can explain the critical effects in circular geometries.
However, it cannot account for the drastically different critical velocities
for rotation and linear flows \cite{channel}.

A superflow of a dilute Bose gas, described by
nonlinear Schr\"odinger equation
\cite{ns},
has been studied recently in different geometries with the use of
direct numerical methods \cite{numerical1}-\cite{numerical2}.
It has been observed that the distinct critical velocity is
linked to the emission of vortices. This velocity turns out to be equal to
the Bogoliubov \cite{bogoliubov} velocity of sound propagation in the gas.
It is in good agreement with the criterion
(\ref{criterion}) since the Dirichlet boundary conditions
(for Bose-gas wave function) are imposed on the channel walls \cite{numerical2}.
This means that the walls, being considered as rigid immovable bodies, 
have no degrees of freedom.
As a consequence, 
the arguments leading to formulas (\ref{criterion}),(\ref{velocity})  can be 
used.

In reality, the channel walls have a large number of degrees of freedom.
Indeed, short- and long-range forces 
between a particle and a surface, 
boundary and interface phonons are well known subjects in the 
surface physics \cite{Gennes},\cite{Luth}.
This means that the (superfluid) Bose-gas is coupled with the channel, in 
which the gas moves. Then special boundary excitations can exist in 
the system of Bose-gas $+$ channel walls because of the coupling between, 
say,
the surface phonons of the walls and the Bogoliubov phonons of the 
Bose-gas.  
Therefore, the Landau criterion in the form (\ref{criterion}) 
cannot be applied; it has to be modified.
To get an analog of it we use the laws of conservation,
taking the (inner) walls' degrees of freedom into consideration.
The superflow can exist, provided the following condition holds:
\begin{equation}
\epsilon (k)>0,\,{\rm in\,the\,channel\,(laboratory)\,reference\,frame},
\label{new}
\end{equation}
where $ \epsilon (k) $ is the energy of {\it any} elementary 
excitation of the
{\it whole}
(gas\,$+$\,walls) system.

Condition (\ref{new}) means that the state of the system can not be changed,
since the occurrence of any number of elementary excitations leads to the
increase of a total energy,
but the latter is prohibited by the low of conservation.
The excess of the momentum is ``absorbed'' by the motion of the center
of mass of the walls. The energy is not actually changed by this motion (in 
the channel reference frame)
because of a large mass of the walls and their zero initial velocity.

\section{Bose-Gas with Boundary Interactions} 

We study the model of the dilute Bose-gas in the channel of the width
$ 2l $ (in $y$-direction) and of infinite length (in $x$- and 
$z$-directions). The
channel walls occupy $ \vert y\vert>l $ part of the space (see Fig. 1).
The general structure of the Hamiltonian is the following:
$$
\hat{H}=H_{\rm gas}( \hat{\psi},\,\hat{\psi}^{\dag})+ 
   {H_{\rm ph} }_{1}(\hat{q}_{1},\,\hat{\pi}_{1})+ 
   {H_{\rm int} }_{1}(\hat{q}_{1},\,\hat{\psi}^{\dag} \hat{\psi}) +
$$  
\begin{equation}
   + {H_{\rm ph} }_{2}(\hat{q}_{2},\,\hat{\pi}_{2})+ 
   {H_{\rm int} }_{2}(\hat{q}_{2},\,\hat{\psi}^{\dag} \hat{\psi}), 
\label{ham1}
\end{equation}     
where 
$\hat{\psi}$ is the Bose-gas field operator, $\hat{q}$ is the 
displacement field operator of a wall, $\hat{\pi}$ is the momentum density 
operator
conjugate to $\hat{q}$, and the indexes 1,2 correspond to the upper and 
lower  part of the channel respectively.
In the model being considered, the Bose-gas Hamiltonian has the following
form:
$$
H_{\rm gas}=\!\int\!\hat{\psi}^{\dag}({\bf 
r})\left(-\frac{\hbar^{2}}{2m}\Delta\right) \hat{\psi}({\bf r})\,d{\bf r} +\,
\int\!\frac{\nu}{2}\delta({\bf r}-{\bf r}')\hat{\psi}^{\dag}({\bf r})
                                 \hat{\psi}^{\dag}({\bf r}')
                                 \hat{\psi}({\bf r}')\hat{\psi}({\bf r})
\,d{\bf r}d{\bf r}',
$$
where $\nu > 0$ is the interparticle interaction constant 
\cite{Huang1}, 
and the wall Hamiltonian can be written as follows 
$$
H_{\rm ph}=\int\! \frac{ \hat{\boldsymbol \pi }^{2}({\bf r})}{2\rho 
} \,+ \,\partial_{j}\hat{q}_{k}({\bf r})\lambda_{jkln}\partial_{l}
\hat{q}_{n}({\bf r})\,d{\bf r}, 
$$
where the tensor $\lambda_{jkln}$  
describes   
the elastic properties of the channel walls. 
  
We derive the excitation spectrum using the techniques of semiclassical 
approximation
(cf. \cite{bogoliubov},\cite{soliton},\cite{Pit},\cite{Fetter}). 
Expanding the field operators
near certain classical solutions,
i.e. $ \hat{\psi}= \psi_0+\delta\hat{\psi} $ 
and 
$ \hat{q}= q_0+\delta\hat{q}$,\,
we represent the Hamiltonian in the form
\begin{equation}
\hat{H}=H_0+\hbar \hat{H }_2+\dots ,
\label{quasi}
\end{equation}
where $ H_0 $ stands for the classical part of $\hat H$.
Note that 
$\psi_{0} \ne 0$ indicates existence 
of a condensate
in the moving Bose-gas, whereas $q_{0} \ne 0$ appears in this model
mainly to satisfy boundary conditions (see below).
The Hamiltonian 
$\hat{H}_2$ in (\ref{quasi}) is bilinear with respect to the field operators.
As a consequence this (semiclassical) Hamiltonian can be reduced to 
the normal form
\begin{equation}
\hat{H }_2=\sum _i \omega_i \hat{b}_i^{\dag} \hat{b}_i+{\rm const},\, 
\left[ \hat{b}_i,\hat{b}_j^{\dag}\right]=\delta_{ij},
\left[\hat{b}_i, \hat{b}_j\right]=0.
\label{normal}
\end{equation}
The quantum Heisenberg and the classical Poisson-Hamilton 
equations of motion for
the field operators (functions)
\begin{equation}
i(d \hat{b}_i/dt)=
\left[\hat{H},\hat{b}_i\right],\quad (d b_i/dt)=\left\{H,b_i\right\},
\label{heisenberg}
\end{equation}
have the same form, if in (\ref{quasi}) we neglect the terms of power in 
$\hbar$ greater than one.
These equations are linear with respect to the field variables.

It follows from (\ref{normal}), (\ref{heisenberg}) 
that the excitation energies $\omega_i$
are equal to the characteristic frequencies of these equations.
Thus, to determine the semiclassical energy spectrum of the system we 
need to 
find the characteristic frequencies of the classical field equations
for  $\delta \psi $ and $\delta  q$
linearized around a proper stationary 
solution $ \psi_0,\, \,q_0 $.
(This means that 
 $\delta \psi$ and $\delta q$, originally of the operator nature, can be  treated 
as a c-number and a real number function respectively). 
Then
the gas state can be characterized by 
the Bose wave function $ \psi({\bf r}, t) = \psi(x,y,t) $, while
the state of the walls is determined by the displacement field
$ {\bf q}({\bf r},t)=(q_x(x,y,t),\,q_y(x,y,t),\,0) $. For simplicity,
the system is asumed to be homogeneous in the z-direction.

\begin{figure}
\begin{center}
\leavevmode
\epsfxsize = 220pt
\epsfysize = 160pt
\epsfbox{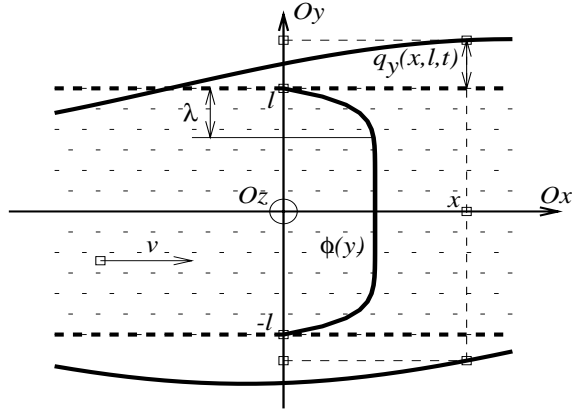}
\end{center}
\caption{The Bose gas moves with the velocity ${\bf v}$ in a channel of the width $2l$.
The profile of the stationary wave function
$ \phi(y)=\vert\psi_0(x,y,t)\vert $ is shown in the center, where $\lambda $ is the  coherence
length.
The top and bottom curves depict the boundaries of  the walls interacting with the gas,
while the horizontal bold dashed lines correspond to the unperturbed boundaries,
and $q_{y}(x, \pm l, t)$ are deviations from  this equilibrium.} 
\end{figure}

The interaction between the wall and gas atoms 
 is given in our model by the sum of hard-core
repulsion and some ``smooth'' potential 
$ {\cal {U} }= {\cal {U}}({\bf r},\,\psi^{*}\psi, \,{\bf q})$. 
The hard-core repulsion makes the walls impenetrable for
 the gas. It follows that the wave
function of the Bose-gas vanishes along the actual 
 boundaries $ y_\pm=\pm l + q_y(x,\pm l,t) $
(see Fig. 1).
\begin{equation}
\psi\left(x,\pm l +q_y(x,\pm l,t),t\right)=0.
\label{1}
\end{equation}
The other boundary conditions, namely
\begin{eqnarray}
& & \sigma_{yy}(x,\pm l,t)={\hbar^2 \over m} \partial_{y}\psi^{*}\,\partial_y\psi(x,y=\pm l,t),
\label{2}\\
& & \sigma_{xy}(x,\pm l,t)=0,
\label{3}
\end{eqnarray}
correspond to the equality of the forces between the solid and gas
on the boundary. In (\ref{2}), (\ref{3})
$\sigma_{ij}$ denotes the wall stress tensor and  $m$ is a mass of the gas atom.
The pressure of the gas (r.h.s. of (\ref{2})) is equal to the normal component of the stress
tensor (l.h.s. of (\ref{2})). Eq. (\ref{3}) follows from the fact that 
the tangent stress vanishes on the boundary.

The wave function $ \psi $ of the repulsive Bose-gas
satisfies the nonlinear Schrodinger equation \cite{ns}
\begin{equation}
\left\{i\hbar \partial_t + {\hbar^2 \over 2m} \Delta -\nu 
\psi^{*}\psi \right \}\,\psi=V({\bf r},\, \partial_{j}q_{k})\,\psi,\,\,\, \vert y 
\vert <l, \label{NS}
\end{equation}
while the wall dynamics obeys the hyperbolic equation for
the displacement field $ {\bf q} $ \cite{el},
\begin{equation}
\rho\partial^2_t q_i=\partial_j \sigma_{ij}-\partial_i W({\bf r},\,\psi^{*}\psi),\quad \vert y \vert >l,
\label{hyp}
\end{equation}
($\rho$ denotes the wall mass density).
The potentials $V$ and $W$ depend on the ``smooth'' part of the gas-wall 
interaction $\cal {U} $; they must vanish if\, $ {\cal {U}}=0 $.
To make our model as simple as possible, we set ${\cal {U}}=0 $ {\it a priori}.
Then, dynamics is described by equations
(\ref{NS}), (\ref{hyp}) with constant coefficients, whereas the hard-core 
interactions fix the boundary conditions (\ref{1})-(\ref{3}).
Note that these conditions imply that the walls have a finite compressibility,
$K=-V(\partial_{V}\,p)_{S} <\infty $. Therefore, even though ${\cal U}=0$,  
\,the wall and Bose-gas excitations 
can be coupled because $(\partial_{y}\psi )\,q_{y}(x, \pm l)$ 
in (\ref{1}) and $\sigma_{yy}\sim K\partial_{y}q_{y}(x, \pm l)$ 
in (\ref{2}) are finite and time dependent.
  
We suppose that the walls are isotropic (this assumption does not affect 
results qualitatively, simplifying our calculations), so that \cite{el}
\begin{equation}
\sigma_{ij}=\rho c_t^2\left(\partial_j q_i+\partial_i q_j+
(\beta-1)\delta_{ij}({\boldsymbol \nabla} {\bf q})\right),\, \, \beta=(c_l^2/c_t^2) -1, 
\end{equation}
where $c_{l}$, $c_{t}$ are longitudinal and transversal sound velocities
respectively.
It is convenient to rescale variables in such a way that the spatial coordinates
are expressed via the coherence length $\lambda$ units while the flow velocity is measured
in terms of the Bogoliubov sound velocity $ c_{\rm B} $ \cite{bogoliubov}:
$$
\lambda={\hbar/\sqrt{\nu\rho_{\rm gas}}},\quad c_{\rm B}={\sqrt{\nu\rho_{\rm gas}}/m}.
$$
Here, $ \rho_{\rm gas} $ stands for the bulk gas density.
Then Eqs. (\ref{NS}),(\ref{hyp}) become
\begin{equation}
\left(i\partial_{t} + (1/2)\Delta - \psi^{*}\psi\right)\psi(x,y,t)=0,
\label{NS1}
\end{equation}
\begin{equation}
\left\{c_t^{-2}\partial^2_t-\Delta-
\beta {\boldsymbol \nabla} ({\boldsymbol \nabla} \cdot)\right\} {\bf 
q}(x,y,t)=0, 
\label{hyp1}
\end{equation} 
where the sound velocity, time and ${\bf q}$ are measured in the units of 
$c_{\rm B}$, $\lambda/c_{\rm B}$ and $\lambda$ respectively.

In the stationary regime, the Bose gas moves uniformly in the 
$x$-direction with the velocity $v$.
As the system is homogeneous in the $x$ and $z$ directions,
$\vert\psi_0({\bf r},t)\vert=\phi(y)$ and ${\bf q}_{0}({\bf r},t)={\bf 
q}_{0}(y)$.
The walls are deformed only in the $y$-direction, since the tangent stress vanishes
in the stationary regime ${\bf q}_{0}(y)=\left (0,\,{q_{0}}_{y}(y),\,0 \right )$.
The corresponding solution of (\ref{NS1}), (\ref{hyp1}) is given by
\begin{eqnarray}
& & \psi_0(x,y,t) = \phi(y){\rm e}^{ivx}
\,{\rm e}^{-i\Omega t},\,\,\,\phi(\pm l)=0,\,\,
\phi'(\pm l)\ne 0,
\label{stationary0} \\
& & {q_0}_{x} = 0,\,\, \,{q_0}_{y}\equiv Q(y)=\pm{\rm const}(y\pm l),
\,(y\rightarrow  \pm l),\,\, Q(\pm l)=0,\,\, Q'(\pm l)\ne 0
\label{stationary1}
\end{eqnarray}
It follows from (\ref{stationary0}), ({\ref{NS1}) that
$ \Omega=\tilde{\mu} + {v^2\over 2} $,\, $\tilde{\mu} = 1$ (this corresponds to the 
value of a chemical potential \,$\mu=\nu\rho_{\rm gas}/m$ at $T=0$),
and $ \phi(y) $ satisfies the following equation \cite{Fetter1}:
\begin{equation}
-{1\over 2}\phi^{\prime\prime}+\phi^3= \phi, \quad \phi(y)=\phi(-y) .
\label{stationary2}
\end{equation}
The parity of $ \phi $ in (\ref{stationary2}) and 
the boundary conditions in (\ref{stationary0})
are obtained from (\ref{1})-(\ref{3}) (see Fig. 1).

We follow the procedure (\ref{quasi})-(\ref{heisenberg})
expanding the field variables around the stationary solution
(\ref{stationary0}), (\ref{stationary1})
\begin{eqnarray}
& & \psi=\Bigl(\phi(y)+\xi(x-vt,y,t)\,\Bigr){\rm e}^{ivx}\,{\rm 
e}^{-i(\tilde{\mu}+ v^{2}/2) 
t},\nonumber\\
& & q_x=\zeta_x(x-vt,y,t),\,q_y=Q(y)+\zeta_y(x-vt,y,t).\nonumber
\end{eqnarray}
Substituting these expansions into (\ref{NS1}),(\ref{hyp1})
we get 
the following linear differential equations for fluctuations
$ \xi,{\boldsymbol \zeta}$, 
\begin{eqnarray}
& & \left\{ i\partial_t+{1\over 2}\Delta+1-2\phi(y)^2\right\}\xi-\phi(y)^2\xi^*=0, \,
\vert y \vert < l,
\label{linear1} \\
& & \left\{c_t^{-2}(\partial_t-v\partial_x)^2-\Delta-\beta{\boldsymbol \nabla}({\boldsymbol
\nabla}\cdot)\right\}{\boldsymbol \zeta}=0,
\quad \vert y \vert >l,
\label{linear2} \\
& & \xi=\xi(x,y,t), \quad {\boldsymbol \zeta} = {\boldsymbol \zeta}(x,y,t),
\nonumber
\end{eqnarray}
written in the reference 
frame moving with the Bose-gas, $ x\,\rightarrow\,x'=x-vt $.

One of the advantages of setting ${\cal {U}}=0$ is the possibility of using
the exact solution of Eq. (\ref{stationary2}) with the boundary conditions
(\ref{stationary0}),\, 
$$
\phi(y)={\rm sn}(y+l,\,\varrho ),
$$
where ${\rm sn}(y,\varrho)$ is the elliptic sine \cite{sine},
the parameter $\varrho$
is chosen to fit the boundary conditions and the following condition
holds
 ${\rm sn}(y,\varrho)\rightarrow
{\rm tanh}(y)$ if $l\rightarrow \infty $. (Notice that the dimensional 
condensate wave function can be written in the form 
\,$\phi_{d}(y)={\rm const}\sqrt{\rho_{\rm {gas}}/m}\,
{\rm {sn}}\bigl(\,(y+l)/\lambda,\,\varrho\bigr)$\,).

Nontrivial excitations can not propagate over the wall region far 
from the boundary.
Indeed, the equation (\ref{linear2}) has the 
constant coefficients and, hence, the 
dispersion law for such excitations coincides with the phonon one
(i.e. corresponding asymptotical solutions describe propagation
of the ordinary sound waves far from the boundaries). 
Therefore, we have to look for the solution of (\ref{linear2}) decreasing
in $ y \to \pm \infty $ directions.
We assume that the experimentally discovered dependence of the critical
velocity on the canal width \cite{putterman}, $v_{c} \sim l^{-n}$, $n 
\simeq 2$, is a hint to search for the special type of
excitations, in which two boundaries of the canal can contribute {\it 
coherently}. 

Such a solution of (\ref{linear1}), (\ref{linear2}) can be written in the form
\begin{eqnarray}
& & \xi=\chi_1(y)\sin(kx-\omega t)+i\chi_2(y)\cos(kx-\omega t),
\,\,\vert y \vert < l,
\label{ansatz1} \\
& & \zeta_x=r_1(y)\cos(kx-\omega t),\,\,\zeta_y=r_2(y)\sin(kx-\omega t),
\,\, \vert y \vert >l
\label{ansatz2}
\end{eqnarray}
with
\begin{equation}
r_i (y) = A_i\exp(-\kappa\vert y \vert)+B_i\exp(-\eta\vert y \vert).
\label{boundary}
\end{equation}

Two exponential terms in (\ref{boundary}) correspond to the different polarizations of the
boundary excitations. The characteristic
values $\kappa =\sqrt{k^2-\omega^2/c_t^2}, \eta =\sqrt{k^2-\omega^2/c_l^2} $
are eigenvalues of the ordinary linear equations
obtained by substitution of  (\ref{ansatz2}) into (\ref{linear2}).
Note that the ansatz (\ref{ansatz1}),\,(\ref{ansatz2}) is equivalent to 
the Bogoliubov $u$-$v$ transformation generalized to a
nonuniform case \cite{bogoliubov},\cite{Pit},\cite{Fetter} and coupling with
the surface phonons:
\begin{eqnarray}
e^{i\mu t}\delta\psi(x,y,t) &=& u_{k}(y)\,{\rm e}^{i(kx-\omega(k) t)} + 
                       v^{*}_{k}(y)\,{\rm e}^{i(-kx+\omega(k) t)}\nonumber\\
\delta {\bf q}(x,y,t) &=& {\bf C}_{k}(y)\,{\rm e}^{i(kx - \omega(k) t)} + 
{\rm c.c.}. \nonumber
\end{eqnarray}
Then the operators $b_{k}^{\dag}$, ($b_{k}$) that create (annihilate) the
boundary excitations (\ref{ansatz1}-\ref{boundary}) in the diagonalized 
Hamiltonian (\ref{normal}) can be represented by the linear combinations of the
Bose-gas field operators, $\delta\hat{\psi}$ and $\delta\hat{\psi}^{\dag}$,
and the displacement field operators, $\delta\hat{q}$ and $\delta\hat{\pi}$.

We linearize the boundary conditions (\ref{1})-(\ref{3}) according to
the method of the semiclassical
approximation (\ref{quasi}),
neglecting the terms of power greater than one in
$ \xi,{\boldsymbol \zeta} $. Together with the proper
solution 
(\ref{ansatz1}),(\ref{ansatz2}) of Eqs.\,(\ref{linear1}),\,(\ref{linear2}),
conditions (\ref{1})-(\ref{3}) determine
values $ A=A(k,\omega)$, $B=B(k,\omega) $ in (\ref{boundary}) and
the boundary conditions for $ \chi_{1,2} $:
\begin{eqnarray}
& & \left({\chi_1^{\prime} \over \chi_1}\right)_{y=\pm l}=
\mp k\,\gamma (z),\,\, \chi_2(\pm l)=0 , \quad  z={\epsilon (k)^2 \over 
(c_t k)^{2}} , \label{conditions}   \\
& & \gamma(z)={{\rho c_t^2/\rho_{\rm gas}c_{\rm B}^2}\over\left(\phi^\prime(l)\right)^2}{1\over
2z}\left(4\sqrt{1-z}-{(2-z)^2\over\sqrt{1-c_t^2z/c_l^2}}\right)\gg 1.
\nonumber
\end{eqnarray}
The value $ \epsilon(k) $ in (\ref{conditions})
$$
\epsilon(k)=\omega(k) - kv
$$
equals the boundary excitation energy in the channel reference system 
(\ref{new}).

It follows from (\ref{ansatz1}) and (\ref{linear1}) that
the variables $ \chi_{1,2}(y) $ and  $\omega$ satisfy the 
following system: \begin{eqnarray}
& & L_1\chi_1(y)+2\omega\chi_2(y)=0,\quad L_1=\partial_y^2-k^2+2-6\phi(y)^2,
\nonumber \\
& & L_2\chi_2(y)+2\omega\chi_1(y)=0,\quad L_2=\partial_y^2-k^2+2-2\phi(y)^2.
\label{system}
\end{eqnarray}
Note that in view of the parity of $ \phi(y) $ (see (\ref{stationary2}))
and the boundary conditions (\ref{conditions}), solutions of 
(\ref{system}) can be either
symmetric or antisymmetric with respect to $ y $.

According to the criterion (\ref{new}), a breakdown of superfluidity occurs  
at such a value $v$ if there exists such a
$ \tilde{k} \neq 0 $ that $ \epsilon(\tilde{k})=0 $. Then
the argument $ z $ of $ \gamma $ in (\ref{conditions}) vanishes and the 
boundary conditions
do not depend explicitly on $ \omega $, i.e. 
 $(\chi_{1}^{\prime}/\chi_{1})\vert _{y=\pm l}=\mp \tilde{k}
\gamma(0)=\mp {\rm const}$. In principle, this fact makes it possible
to calculate $v_{c}$ without finding any final expression of $\omega(k)$.
Indeed, one has to solve Eqs. (\ref{system}) with $\omega=\tilde{k}v_{c}$
and fixed boundary conditions. 

In this simple model, however, it is possible to calculate 
the dispersion relation $\omega (k)$, at least in the $k \rightarrow 0$ 
limit \cite{Loutsenko11}.    
We look for the symmetric solution of (\ref{system}),
expanding $ \chi_{1,2}(y) $ in powers of $ \omega^2 $:
$$
\chi_1=\chi_1^{(0)}+\omega^2\chi_1^{(1)}+\dots,\quad
\chi_2=\omega(\chi_2^{(0)}+\omega^2\chi_2^{(1)}+\dots).
$$
The functions $ \chi_{1,2}^{(i)} $ symmetric in $ y $ satisfy the following recurrence
relations
\begin{eqnarray}
& & L_1\chi_1^{(0)}=0 ,\, \chi_1^{(0)}(\pm l)=1 ,\nonumber \\
& & L_1\chi_1^{(i)}+\chi_2^{(i-1)}=0,\, \chi_1^{(i)}(\pm l)=0,\,
i=1,2,\dots \nonumber \\
& & L_2\chi_2^{(i)}+\chi_1^{(i)}=0,\, \chi_2^{(i)}(\pm l)=0,\,
i=0,1,2,\dots.
\label{rec}
\end{eqnarray}
The analytic study of (\ref{rec}) seems to be difficult.
Instead we proceed numerically \cite{fortran};
we obtain the eigenvalues $ \omega=\omega(k) $ by solving (\ref{rec}) 
recursively
and imposing the boundary conditions (\ref{conditions}) on $ \chi $.

The result reads
\begin{equation}
\omega(k) = \alpha\sqrt{(\gamma+\delta) k^3}, \quad {\rm as} \quad k \to 0, 
\label{spectrum}
\end{equation}
where $ \alpha >0 $ and $ \delta >0 $ are some bounded functions of $ l $,
$ \gamma=\gamma(0) \simeq 
\rho c_{t}^{2}/\rho_{\rm gas}c_{\rm B}^{2}$.      
The dependence of $ \log(\omega(k)) $ on  $ \log(k) $ is shown on Fig.  2.
Note that only the solutions with finite $\gamma >  1$  have  the physical 
meaning. 
For ``conventional'' systems, such as a dilute Bose gas inside a solid container, 
the value of $\gamma(0)$ is very large, $\gamma(0)\simeq 10^{7\sim 8}$ or even more.
Moreover, the validity of the long-wavelength\,/\,low-energy approximation  
implies $k\gamma < 1$ and $\omega \ll 1$, and the relevant wavelengths are unphysically
huge.  
However, for the ``soft matter'' substances with flexible walls, $\gamma$
can be of the order of $10^{2 \sim 3}$ and the distance between the walls 
can be $2l > \lambda$. Then, beginning from the wavelengths of the order of $(10^{2.5 \sim
3})\lambda $, we are within the `$k\rightarrow 0$' limit 
and $\omega(k) < 10^{-3}\tilde{\mu}$. 
(On Fig. 2, we present also the curves with $\gamma < 1 $  
because of   
similarity between our result and the dispersion relation of capillary 
waves on the interface between 
liquid and gaseous He, the so-called ``ripplons'' \cite{ripplons}.)   
If the channel walls were rigid and incompressible,  
that corresponds to $\gamma =
\infty $  in (\ref{spectrum}),
the inhomogeneous surface excitations introduced in this
study just do not exist as a well-defined object.
   
It is easy to conclude that
the  semiclassical critical velocity is {\it zero} in this {\it model}, 
since for any $ v $ there exist
$ k\neq 0 $ such that $ \epsilon(k)=\omega(k)-kv<0 $. 
This means that the model, in which all the gas\,--\,boundary interactions 
are reduced to the hard-core repulsion, predicts (in the semiclassical 
(Bogoliubov)
approximation) an instability of the Bose-gas current state in relation
to occurrence of boundary excitations.     
However, whether the damping of superflow can happen via the
energy transfer from the 3D condensate to the boundary localized modes
is an open question, which  cannot be answered in the framework of 
models with the superfluid density $\rho_{s}=\rho_{\rm gas}$, $T=0$.   
More sophisticated models of superfluidity, such as the two-fluid 
hydrodynamics \cite{putterman}, should be used to describe the kinetics 
of damping.      
   
\begin{figure}
\begin{center}
\leavevmode
\epsfxsize = 220pt
\epsfysize = 220pt
\epsfbox{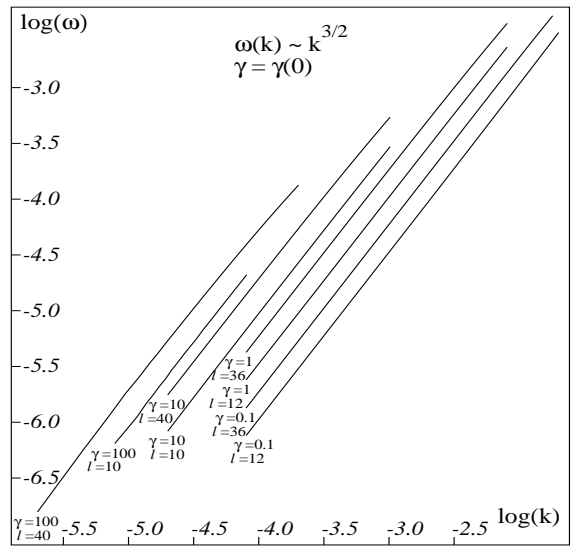}
\end{center}
\caption{The low-momentum spectrum $ \omega(k) $ of boundary excitations.
The canal width $l$ is measured in the coherence length units;\, 
$\gamma \simeq \rho c_{t}^{2}/\rho_{\rm gas}c_{\rm B}^{2}$.      }
\end{figure}

\section{Discussion}

In our opinion, the benefit from finding such a dispersion relation 
(see Fig. 2) in the framework of the proposed
 oversimplified  model 
is the possibility of concluding that
the boundary interactions can play a key role in the
superfluidity, namely, by reducing substantially the critical velocity of 
the superflow. If that is the case, it is reasonable to generalize 
our model to make it more 
realistic. 
   
Recall that we have neglected the ``smooth'' part ${\cal {U} }$ 
of the gas-wall interactions.
The models with repulsive ${\cal { U}} $ seems to be qualitatively 
similar to the one under
consideration. 
We believe that they also lead to the zero {\it semiclassical} critical 
velocity, since nothing can apparently
prevent the energy transfer from the gas flow to the walls
(in this approximation).
It is important to note that even if the higher quantum corrections
yield nonzero critical velocity,
the latter will be of much smaller value than the (semiclassical) 
Bogoliubov-Landau
velocity found by the numerical simulations based on the nonlinear 
Schr\"odinger equation \cite{numerical2}.

The situation might be different if the ``smooth'' part $\cal {U} $ of the 
interaction
between the gas and wall atoms were attractive. 
For example, we can take the Hamiltonian of gas-wall interaction in the 
Deformation Potential form
\begin{equation}
\hat{H}_{\rm int}=\int \sigma ({\bf r},{\bf r}')\, 
  \boldsymbol{\nabla}\hat{{\bf q}}({\bf r})\, \hat{\psi}^{\dag}\hat{\psi} 
({\bf r}')\,d{\bf r}d{\bf r}', 
\end{equation}
where ${\bf r}$ and ${\bf r}'$ change in the wall area ($\vert y \vert > l $) and 
in the channel area ($\vert y' \vert < l $) respectively,  and the function
$\sigma ({\bf r},{\bf r}')$ describes  atom-lattice interaction.
(At least two new parameters, which control the ``smooth'' part 
of gas-wall interaction, have to appear in the model with ${\cal {U}} \ne 0 $:
one, some characteristic value of energy, and, two, a length scale).
Then the equations for the classical parts of $\hat{\psi}$ and $\hat{q}$,
namely $\phi(y)$ and ${q_{0}}_{y}$, 
(see Eqs. (\ref{stationary0}),\,(\ref{stationary1})\,), can be written
as follows:
\begin{equation} 
\left ( \mu +  \frac{\hbar^{2}}{2m} \partial_{y}^{2}
   -\, \nu \phi^{2}(y) \right ) \phi(y) = 
 \left (\int \sigma({\bf r}',{\bf r}) \partial_{y'} Q(y')\,d{\bf r}' \right ) 
\phi (y),
\label{cond1}
\end{equation} 
\begin{equation}
-c_{l}^{2}\,\partial_{y}^{2} Q(y) = \rho^{-1} \partial_{y}
 \int \sigma ({\bf r}, {\bf r}')\,\phi^{2}(y')\,d{\bf r}'.  
\label{uuu}
\end{equation}
After the exclusion of $Q(y)$ from Eq. (\ref{cond1}), it can be rewritten in 
the following form: 
\begin{equation}
 \left (- \frac{\hbar^{2}}{2m} \partial_{y}^{2}
+ \Lambda \,U({\bf r})+ 
\nu \phi^{2}(y) - \!\int U_{\rm eff}({\bf r}, {\bf 
r}') \phi^{2}(y')d{\bf r}' \right) \phi(y) = \mu\phi(y)
\label{cond2} 
\end{equation}
where $U_{\rm eff}({\bf r}, {\bf r}')=\int\! \sigma({\bf r}'', {\bf r})
\sigma({\bf r}'',{\bf r}') d{\bf r}''/\rho c_{l}^{2}$,
\,$U({\bf r})=\int\! \sigma({\bf r}', {\bf r}) d{\bf r}'$, and $\Lambda = {\rm const}$
is defined from the boundary condition (\ref{2}).
 
It is easy to see from the structure of 
(\ref{cond2}) that 
the atom-lattice  interaction $\sigma({\bf r}, {\bf r}')$
(when exceeding a certain magnitude) can induce an attraction 
between the gas atoms in
the boundary region. Indeed, there can exist such a scale of
$\vert {\bf r} - {\bf r}' \vert $ that    
\,$\nu\delta({\bf r}-{\bf r}') - U_{\rm eff}({\bf r}, {\bf r}')\,< 0$.
In that case, one would expect essential changes in 
the spectrum of boundary excitations. The study of exciton superfluidity
\cite{exciton},\cite{miscellaneous} hints such a possibility.

This study reveals a mechanism of the exciton-exciton attraction induced 
by the lattice effects.
The exciton gas falls into the soliton-like state $\psi_{0}$, $q_{0}$,
when the exciton-phonon
coupling constant exceeds a certain value, or, equivalently, 
the velocity $v$ exceeds some critical value. The excitation spectrum  
in an exciton branch has a gap, and   
the Landau critical velocity, calculated for this type of excitations, 
is given by
\begin{equation}
v_c \simeq \hbar/mL,
\label{critical}
\end{equation}
where $ L $ denotes the characteristic size of the soliton.
We can adapt the similar considerations in our model, 
replacing the exciton-phonon by
the gas-wall interactions.
The critical velocity can be given by a formula similar to (\ref{critical})
with $ L \simeq l $, 
provided $ l $ is the only macroscopical length in 
the theory. 

In this article we did not consider the influence of  
the long-range van der Waals forces between the walls
and gas atoms on the stability of superfluid flow.
Although the attractive part of these forces originates from interaction 
between the electron shells of the particles  
\cite{Celli}, it can be included in our model in the 
form of a static van der Waals potential appearing in the r.h.s. of Eq. 
(\ref{NS}).  
Such an external potential being localized near the 
boundaries can be stronger than the effective potential
$\Lambda U(y)$ in (\ref{cond2}) and therefore can change  
the properties of the condensate wave function, the boundary 
excitations, etc..  

\section{Conclusions}

In conclusion, our simple model manifests one of the possible microscopic 
mechanisms for dissipation processes in the quantum Hamiltonian system of 
coupled Bose-gas and channel walls. 
We show that the dissipation 
can also be caused by creation of boundary excitations in this system. 
Although, in the semiclassical approximation, 
this process is not prohibited at any velocity of the moving repulsive 
Bose-gas, the higher quantum corrections to the self-energy part of the 
boundary excitations
may be essential to obtain the nonzero value of $v_{c}$ in the theory
of the dilute Bose-gas
with gas-wall interaction.  
On the other hand, more rigorous consideration  
in the framework of the semiclassical approximation 
should involve solutions of
(\ref{1})-(\ref{hyp}) with ${\cal {U}} \neq 0 $, where the attractive part
of gas-wall interaction is taken into account.
Such solutions can be 
still represented in the form (\ref{ansatz1}), (\ref{ansatz2}), though
equations (\ref{stationary2}), (\ref{linear1}), (\ref{linear2}) would become 
integral. A direct numerical study of the flow dynamics  
would be also useful.

\section*{Acknowledgments}
One of the authors (D.R.) is very grateful to Y.~Berest for valuable remarks
and to I.~Loutsenko for stimulating discussions and critical reading of the
manuscript.


\begin{thebibliography}{45}
\bibitem{fortran}
We solve a difference analog of this system by the Gauss
method using a FORTRAN program. The calculational error does not exceed the one of
the semiclassical approximation.
\end{thebibliography}
\end{document}